\documentclass[twocolumn,showpacs,prb,amsmath,amssymb,floatfix]{revtex4-1}
\pdfoutput=1
\usepackage{amsmath,amssymb}
\usepackage{bm}
\usepackage{graphicx}
\usepackage{psfrag}
\usepackage{color}
\usepackage{hyperref}
\newcommand{\bd}{\bm}

\newcommand{\nspace}{
\vspace{-.3cm}
}
\newcommand{\sect}[1]{\section{#1}}
\begin{document}

\title{Time-dependent spin-wave theory}

\author{Andreas R\"{u}ckriegel, Andreas Kreisel,  and Peter Kopietz}

\affiliation{Institut f\"{u}r Theoretische Physik, Universit\"{a}t
  Frankfurt,  Max-von-Laue Strasse 1, 60438 Frankfurt, Germany}

\date{February 1, 2012}

 \begin{abstract}

We generalize the spin-wave expansion in powers of 
the inverse spin to time-dependent quantum spin models describing
rotating magnets or magnets in time-dependent external fields.
We show that in these cases,
the spin operators should be projected onto
properly defined rotating reference frames
before the spin components are bosonized using
the Holstein-Primakoff transformation.
As a first application of our approach, we calculate the
reorganization of the magnetic state 
due to Bose-Einstein condensation
of magnons in the magnetic insulator yttrium-iron garnet; we predict a characteristic dip
in the magnetization which should be measurable in experiments.

\end{abstract}


\maketitle

\sect{Introduction}

At low temperatures the static and dynamic properties of magnets
are often determined by spin-wave excitations, which are bosonic
quasiparticles in a magnetically ordered state.
The theory of spin waves \cite{Akhiezer68} has been extremely successful to
explain experimental data for a great variety of magnets.
The basic assumption is
that the thermal and quantum fluctuations 
are sufficiently small, so that one can expand in fluctuations around
the classical ground-state configuration. 
The first step in the spin-wave expansion is therefore
the determination of the 
spin configuration in the classical limit,
where the spin operators are treated as classical vectors.
Deviations
from the classical limit can then be obtained by 
projecting the spin operators onto a basis which matches
the direction defined by the classical spin configuration, and then
bosonizing the spin components
using the  Holstein-Primakoff transformation. \cite{Akhiezer68}
Assuming that the spin quantum number $S$ is large,
one can then calculate fluctuation corrections
perturbatively  in powers of $1/S$.

It is not obvious how to generalize this strategy to 
explicitly time-dependent spin Hamiltonians, because in this case energy is not conserved
and the proper basis for setting up the spin-wave expansion may not 
be determined by minimizing the classical ground-state energy.
At the first sight one can avoid this problem
by simply projecting the spin operators
onto a fixed (laboratory) coordinate system and then introducing 
Holstein-Primakoff bosons as usual.
However, as will be demonstrated below, this strategy 
is not suitable to describe a possible dynamic reorganization
of the magnetic state.
Moreover, in the laboratory basis it is often very cumbersome
(and in practice impossible) to take into account the dominant fluctuation effects.
In this work, we shall develop the general framework to
set up a proper $1/S$ expansion out of equilibrium and then
use our method to calculate the
magnetization dynamics of a simplified spin model
for the pumped magnon gas in the magnetic insulator yttrium-iron garnet 
(YIG),\cite{Cherepanov93,Melkov96} where parametric resonance
and Bose-Einstein condensation (BEC)
of magnons has recently been observed. \cite{Demokritov06}

\sect{Spin wave approach}

\label{sec:spinwaves}
\subsection{Spin-wave expansion in equilibrium}
To explain the basic principles of the time-dependent  spin-wave expansion,
we first  consider a Heisenberg ferromagnet
in a time-dependent magnetic field, 
\begin{align}
 {\cal{H}} ( t )   & =   - \frac{1}{2} \sum_{ i j  } J_{ij} {\bd{S}}_i \cdot {\bd{S}}_j
 - \sum_i {\bd{h}}_i (t ) \cdot {\bd{S}}_i,
 \label{eq:Hdefh}
 \end{align}
where the sums are over the $N$ sites of a cubic lattice,
and $\bd{S}_i$ are quantum mechanical spin operators
localized at the lattice sites $\bd{R}_i$.
The spins interact via exchange couplings $J_{ij}$ and are exposed
to an external space- and time-dependent magnetic field $\bd{h}_i ( t )$
which we measure in units of energy.
Assuming that $\bd{h}_i ( t )$ is sufficiently large, 
the nonequilibrium expectation values
$\langle \bd{S}_i ( t ) \rangle$ are finite so that the time-dependent
unit vectors
 $\bd{\hat{m}}_i ( t ) = \langle \bd{S}_i (t ) \rangle /  | \langle \bd{S}_i ( t ) \rangle |$
in the direction of the local magnetic moments are well defined.
If the time-dependence of the external field is sufficiently slow,
we may use the \textit{adiabatic approximation} to determine $\bd{\hat{m}}_i ( t )$.
In this case we may set up the spin-wave expansion 
as in equilibrium \cite{Schuetz03} by 
projecting  the spin operators onto a time-dependent basis
$\{ \bd{e}_i^{(1)} (t ) , \bd{e}_i^{(2)} (t ) , \hat{\bd{m}}_i (t ) \}$, 
where  $\bd{e}_i^{(1)} (t ) $ and $ \bd{e}_i^{(2)} (t )$ are time-dependent unit vectors
orthogonal to  $\hat{\bd{m}}_i (t )$. 
The directions $\hat{\bd{m}}_i (t )$  are determined by a time-dependent extension of the
static minimization condition of
the classical ground-state energy,~\cite{Schuetz03}
\begin{equation}
  \hat{\bd{m}}_i ( t )
  \times
  \Big[
  {\bd{h}}_i ( t )  + S \sum_j  J_{ij}   \hat{\bd{m}}_j ( t ) 
  \Big] = 0.
  \label{eq:classicalt}
\end{equation}
We then expand the spin operators as
$ {\bd{S}}_i =  S^{\parallel}_i \hat{\bd{m}}_i  +
 \frac{1}{2} [ {S}_i^{+} {\bd{e}}^{-}_i     +  {S}_i^{-} {\bd{e}}^{+}_i   ]     $, where
${\bd{e}}^{\pm}_i  =
  {\bd{e}}^{(1)}_i  \pm i  {\bd{e}}^{(2)}_i $.
Finally,  we express the spin components
in terms of canonical boson operators $a_i$
using the Holstein-Primakoff transformation,~\cite{Akhiezer68}
 $ S_i^{\parallel}  =  S- a^{\dagger}_i a_i$, 
 ${S}_i^+ =  ( S_i^-)^{\dagger}  = [  2 S  - a^{\dagger}_i a_i ]^{1/2} a_i$.
For large $S$ the square roots can be expanded and the
interactions between spin-waves can be
taken into account by means of a systematic expansion in powers $1/S$.

\subsection{Spin waves in the adiabatic basis}
\begin{figure}[tb]    
   \centering
  \includegraphics[width=0.3\textwidth]{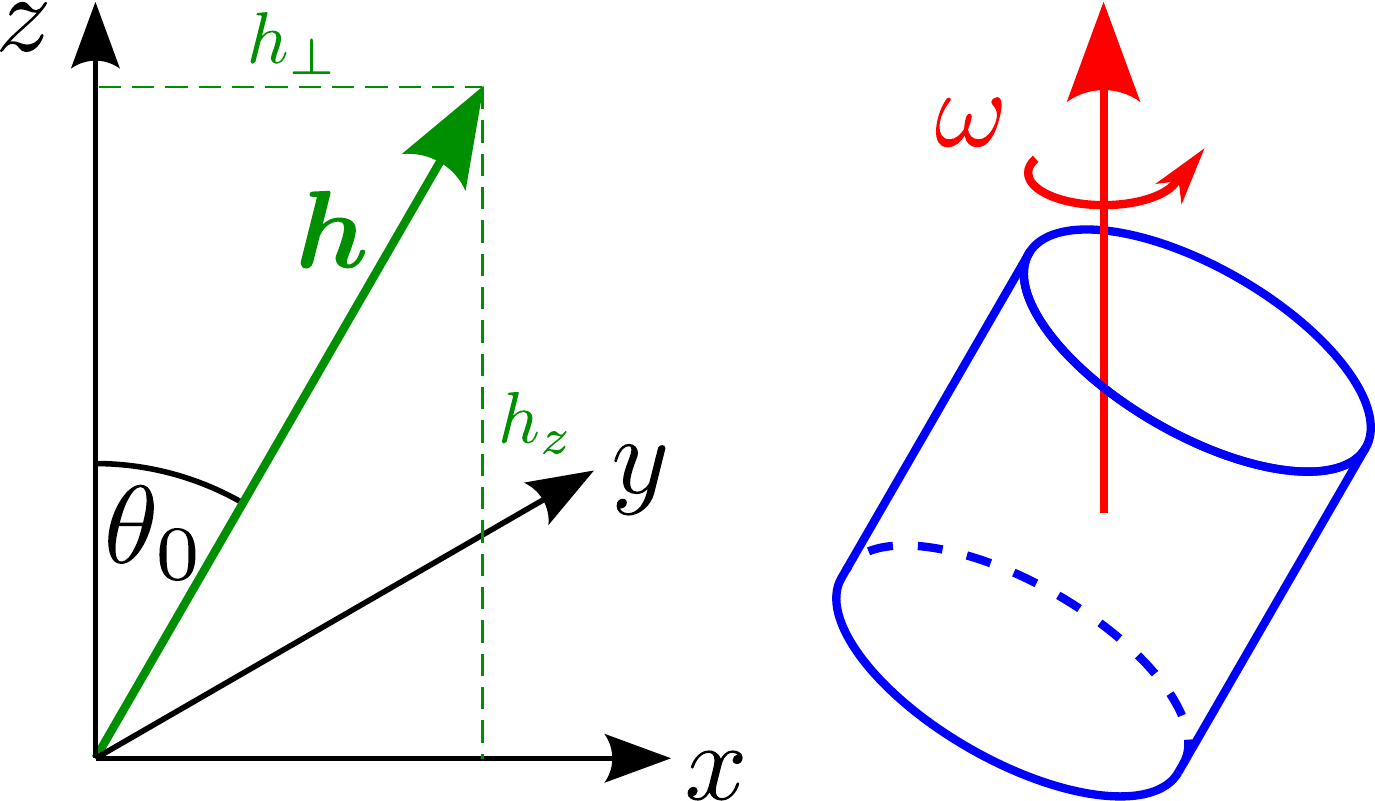}\nspace
  \caption{%
(Color online)
Rotating magnet in a constant magnetic field $\bd{h}$ forming
an angle $\theta_0$ with  the rotation axis $\hat{\bd{z}}$ that is
described by the time dependent Hamiltonian (\ref{eq:Hdefh}).
The system is represented by a cylinder which is actively rotated 
with the constant angular velocity $\omega$
around an axis which is not parallel
to the external magnetic field.
The cylindrical shape and the orientation of the sample reflects the symmetry
of the system at time $t=0$ where the system is rotationally invariant
around the direction of the fixed magnetic field.
}
    \label{fig:heisrot}
  \end{figure}

It turns out, however, that this approach is only useful 
in the adiabatic limit where the rate of change of the external field is small compared
with $ | \bd{h}_i ( t )|$. To see this, 
consider the special case of a homogeneous field $\bd{h}_i ( t ) = \bd{h} ( t )$ which 
rotates clockwise with frequency $\omega$ around the $z$ axis,
\begin{equation}
\bd{h} ( t ) =   h_{\bot} [  \cos ( \omega t )  \hat{\bd{x}} -
\sin ( \omega t )  \hat{\bd{y}}  ] + h_z \hat{\bd{z}}\;, 
\end{equation}
where
$\hat{\bd{x}}$, $\hat{\bd{y}}$ and $\hat{\bd{z}}$ are unit vectors in the
directions of three orthogonal axes of the laboratory. 
By writing~\cite{footnoterot} 
$ \bd{h} ( t ) \cdot \bd{S}_i = \bd{h} ( 0 ) \cdot e^{ \omega t \hat{\bd{z}} \times }
 \bd{S}_i$, we see that Eq.~(\ref{eq:Hdefh}) can alternatively be 
interpreted as the Hamiltonian of a 
magnet which rotates counter-clockwise
with angular velocity $\omega$ around an axis $\hat{\bd{z}}$ which is not parallel
to the field, as shown in Fig.~\ref{fig:heisrot}.
In  adiabatic approximation the magnetization points into the direction of the
magnetic field, as can be easily seen from Eq.~(\ref{eq:classicalt}).
Within linear spin-wave theory we obtain the Hamiltonian
\begin{equation}
 {\cal{H}} \approx \sum_{\bd k} E_{\bd k}^{\rm ad} a_{\bd k}^\dagger a_{\bd k}\;,
\end{equation}
where the ground-state energy has been dropped. Note that the dispersion
$E_{\bd k}^{\rm ad}=\epsilon_{\bd k}+h$ is the sum of the zero-field magnon dispersion
$\epsilon_{\bd{k}} = S ( J_0 - J_{\bd{k}})$ (where $J_{\bd{k}}$ is the
Fourier transform of the exchange couplings $J_{ij}$) and the absolute value 
$h =( h_{\bot}^2 + h_{z}^2 )^{1/2}$ of the magnetic field.
Assuming that at time $t=0$ the system is in thermal equilibrium at inverse temperature $\beta=1/T$,
we find that in adiabatic approximation the time-dependent magnetization 
$ \bd{M} ( t ) = \frac{1}{N} \sum_i \langle \bd{S}_i ( t ) \rangle$
is to linear order in spin-wave theory given by 
 \begin{equation}
 \bd{M}_{\rm ad} (t ) =
 M_{h } \hat{\bd{m}} ( t ) \;,
 \label{eq:Mad}
 \end{equation}
with the magnitude of the magnetization
\begin{equation}
 M_h  = S - \frac{1}{N} \sum_{\bd{k}} \frac{1}{e^{ \beta E_{\bd k}^{\rm ad}  } -1 }\;,
\end{equation}
and its direction
$\hat{\bd{m}} ( t ) =    \sin \theta_0 [  \cos ( \omega t )  \hat{\bd{x}} -
\sin ( \omega t )  \hat{\bd{y}}  ] + \cos \theta_0  \hat{\bd{z}}$.
Here $\theta_0$ is the angle between the magnetic field and the
rotation axis, i.e.
\begin{equation}
\cos \theta_0 = \frac{h_z}{h}
\end{equation}
as shown in Fig.~\ref{fig:heisrot}.

\subsection{Perturbation theory in the laboratory basis}
To see that Eq.~(\ref{eq:Mad})  is only valid for
$ |\omega | \ll h$, let us repeat the calculation of the magnetization
in a \textit{perturbative approach}.
To set up the spin-wave expansion we write our Hamiltonian
 \begin{equation} 
{\cal{H}} ( t ) = {\cal{H}}_z
 + {\cal{V}} ( t )
\label{eq:Hlab}
 \end{equation}
as a sum of the time independent part
\begin{align}
 {\cal{H}}_z   & =   - \frac{1}{2} \sum_{ i j  } J_{ij} {\bd{S}}_i \cdot {\bd{S}}_j
 -  h_z \sum_i S^z_i,
 \label{eq:H0def}
 \end{align}
and the time dependent perturbation
 \begin{align}
  {\cal{V}} ( t ) & =  - h_{\bot} \sum_i \left[ \cos ( \omega t ) S^x_i 
 - \sin ( \omega t ) S^y_i \right]
. \label{eq:H1def}
 \end{align}
We now project the spin operators onto the fixed laboratory basis.
This strategy is usually adopted 
to discuss parametric resonance of magnons 
\cite{Suhl57,Schloemann63,Zakharov70} and has recently been used in Ref.~[\onlinecite{Nakata11}]
to calculate the nonequilibrium dynamics of magnons in a related spin model.
After expressing the Hamiltonian (\ref{eq:Hlab}) in terms of
laboratory-frame Holstein-Primakoff bosons $b_i$ and
transforming to momentum space, $b_i = N^{-1/2} \sum_{\bd{k}} e^{ i {\bd{k}} \cdot
 {\bd{R}}_i } b_{\bd{k}}$,
the Hamiltonian reads in linear spin-wave theory
\begin{equation}
 {{\cal H}_z} \approx \sum_{\bd k} E_{\bd k}^{\rm lab} b_{\bd k}^\dagger b_{\bd k}^{\phantom \dagger}\;.
\end{equation}
The dispersion $E_{\bd k}^{\rm lab}=\epsilon_{\bd k}+h_z$ now contains the static
part of the magnetic field. The time-dependent perturbation Eq.~(\ref{eq:H1def})
can be written as
\begin{align}
 {\cal{V}} ( t ) & =- \frac{h_{\bot}}{2} \sqrt{2S} \sqrt{N} \left[ e^{  i \omega t } b_{\bd{k} =0}^{\phantom \dagger}
 +  e^{ - i \omega t } b^{\dagger}_{\bd{k}=0} \right].
 \end{align}
Since the boson Hamiltonian contains linear terms, the laboratory boson operators
$b_{\bd k=0}^\dagger$ and $b_{\bd k=0}$ have finite expectation values, thus condense.
The dynamics of these expectation values, as well as the time dependence
of the magnon distribution function $\langle b^{\dagger}_{\bd{k}} ( t )
 b_{\bd{k}} ( t ) \rangle$ can be easily obtained within linear spin-wave theory
by solving the Heisenberg equations of motion. With appropriate initial conditions
we obtain for the time evolution of the magnetization,
\begin{align}
 \bd{M}_{\rm lab} ( t ) & =  \frac{h_{\bot} S}{ h_z - \omega} 
 \left[  \cos ( \omega t ) \hat{\bd{x}} - \sin ( \omega t ) \hat{\bd{y}}
 \right] + M_{h_z}  \hat{\bd{z}}.
 \label{eq:Mlab}
 \end{align}
Formally the perturbation has been carried out as an expansion in powers of $h_\perp/h_z$,
but we will see later that it is essentially an expansion in powers of $h_\perp/(h_z-\omega)$.
An important difference to the 
adiabatic result (\ref{eq:Mad}) is the singularity for $\omega \rightarrow h_z$,
which is of course unphysical
and one would need a resummation to all orders in $1/S$ to resolve this.
Using a similar approach, such a singularity
has also been found in
Ref.~[\onlinecite{Nakata11}] for a slightly different model.
Although for $ | \omega  - h_z | \lesssim h_{\bot}$ perturbation theory in the laboratory frame breaks down, Eq.~(\ref{eq:Mlab}) indicates that
both the  adiabatic approximation and the perturbative approach in the laboratory frame
have serious limitations: while the adiabatic basis
is restricted to slowly varying external field and misses possible dynamic instabilities,
in the laboratory basis one generates unphysical singularities in linear spin-wave theory, indicating that important fluctuation effects have been neglected.

\sect{Spin waves in the proper rotating reference frame}
\label{sec:canonical}

We now develop a time-dependent generalization of the spin-wave expansion
which neither suffers from the limitations of the adiabatic approximation nor
exhibits the pathologies of the perturbative approach
in the laboratory frame.
Our theory is guided by the following two insights:
(i) the spin operators should be bosonized  in a \textit{proper rotating basis}
whose third axis $\hat{\bd{m}}_i ( t )$ matches the direction of the true nonequilibrium
expectation value $\langle \bd{S}_i (t ) \rangle$ and (ii) the proper rotating basis
in general does not agree with  
the adiabatic basis defined in Eq.~(\ref{eq:classicalt}).

To construct the proper rotating basis, 
consider  the unitary time-evolution operator ${\cal{U}} ( t )$ of
some  arbitrary time-dependent spin Hamiltonian ${\cal{H}} ( t )$, which 
satisfies the operator equation 
 \begin{equation}
 i \partial_t {\cal{U}} ( t ) =  {\cal{H}} ( t ) {\cal{U}} ( t )\;.
\end{equation}
Making the factorization ansatz
\begin{equation}
 {\cal{U}} ( t ) =  {\cal{U}}_0 ( t ) \tilde{\cal{U}} ( t )
\end{equation}
with some suitable $ {\cal{U}}_0 ( t )$, we find that
 $\tilde{\cal{U}} ( t )$ satisfies
$
 i \partial_t \tilde{\cal{U}} ( t ) =  \tilde{\cal{H}} ( t ) \tilde{\cal{U}} ( t ),
 $
with the effective Hamiltonian
\begin{equation}
 \tilde{\cal{H}} ( t ) =  \tilde{\cal{H}}_A (t )  +\tilde{\cal{H}}_B  ( t ) \;,
\end{equation}
where
 \begin{equation}
 \tilde{\cal{H}}_A (t )  ={\cal{U}}^{\dagger}_0 ( t ) {\cal{H}} ( t )  {\cal{U}}_0 ( t )
\end{equation}
corresponds to the adiabatic approximation, while 
\begin{equation}
 \tilde{{\cal{H}}}_B ( t ) = - i
  {\cal{U}}^{\dagger}_0 ( t )  \partial_t   {\cal{U}}_0 ( t )
\label{eq:HBdef}
\end{equation}
contains all corrections to the adiabatic approximation, including
possible  Berry phases. \cite{Berry84,Bohm03}
We now choose   ${\cal{U}}_0 ( t )$ such that for each lattice site
it rotates the 
$z$ axis of the laboratory to an axis in the direction $\hat{\bd{m}}_i ( t )$
of  the true local magnetization.
This is achieved by setting
 \begin{equation}
 {\cal{U}}_0 ( t )= e^{ - i \sum_i \bd{\alpha}_i ( t ) \cdot {\bd{S}}_i }\;,
 \end{equation}
with suitable rotation vectors $\bd{\alpha}_i ( t ) = \alpha_i ( t ) \hat{\bd{\alpha}}_i ( t )$,
where $\alpha_i ( t )$ is the rotation angle and $\hat{\bd{\alpha}}_i ( t )$ is a unit vector
in the direction of the rotation axis.
The rotated spin operators
can then be written as \cite{footnoterot}
\begin{equation}\tilde{\bd{S}}_i (t )  = 
 e^{ i \bd{\alpha}_i (t) \cdot {\bd{S}}_i }
 \bd{S}_i e^{ - i \bd{\alpha}_i (t ) \cdot {\bd{S}}_i }  =
 e^{ \bd{\alpha}_i (t ) \times } \bd{S}_i\;.
\label{eq:Srot}
\end{equation}
To calculate the corresponding Berry-phase contribution $\tilde{\cal{H}}_B ( t )$
to the effective Hamiltonian in the rotating reference frame,
we use  Feynman's \cite{Feynman51}  representation
\begin{equation}
\frac{d}{dt} e^A = \int_0^1 d \lambda e^{\lambda A} \frac{d A}{dt} e^{ (1-\lambda) A}
\label{eq:Fey}
\end{equation}
of the time derivative of the exponential of an operator $A$
which does not necessarily commute with its time derivative $dA/dt$.
It is convenient to  decompose a general rotation 
into three successive
rotations parametrized by the usual Euler angles 
$\varphi$, $\theta$ and $\psi$ as follows,
\begin{equation}
e^{ \bd{\alpha}_i (t ) \times } = 
 e^{ {\bd{\psi}}_i (t ) \times }
e^{  {\bd{\theta}}_i (t ) \times }
e^{  {\bd{\varphi}}_i (t ) \times }\;,
\end{equation}
where the rotation vectors are 
 $
\bd{\varphi}_i ( t )  =  \varphi_i ( t ) \hat{\bd{z}}$,
$ \bd{\theta}_i ( t )  =  \theta_i ( t ) \hat{\bd{\theta}}_i ( t )$, and
$\bd{\psi}_i ( t )  =  \psi_i ( t ) \hat{\bd{m}}_i ( t ) $. \cite{McCauley97}
Explicitly, the direction of the nutation vector 
$\bd{\theta}_i$ is
 $
 \hat{\bd{\theta}}_i ( t )  = 
 \frac{ \hat{\bd{z}} \times \hat{\bd{m}}_i ( t ) }{
|  \hat{\bd{z}} \times \hat{\bd{m}}_i ( t ) | }$.
To define the spin waves in the proper rotating basis,
we expand the rotated spin operators $\tilde{\bd{S}}_i$ defined
in Eq.~(\ref{eq:Srot}) in the time-dependent right-handed basis formed by the 
following three unit vectors:
  \begin{subequations}
 \begin{eqnarray}
 \tilde{ \bd{e} }^{(1)}_i (t) & = & \cos \psi_i (t) \hat{\bd{\theta}}_i (t )
 + \sin \psi_i (t) \hat{\bd{m}}_i (t ) \times \hat{\bd{\theta}}_i (t ) ,
 \hspace{7mm}
 \\
 \tilde{ \bd{e} }^{(2)}_i (t) & = & - \sin \psi_i (t) \hat{\bd{\theta}}_i (t )
 + \cos \psi_i (t) \hat{\bd{m}}_i (t ) \times \hat{\bd{\theta}}_i (t ),
 \hspace{7mm}
 \end{eqnarray}
 \end{subequations}
and $ \hat{\bd{m}}_i  (t) $. The corresponding spin components are defined by
 \begin{equation}
 \tilde{\bd{S}}_i ( t ) =   
 \tilde{S}^{(1)}_i \tilde{ \bd{e} }^{(1)}_i (t)
+  \tilde{S}^{(2)}_i  \tilde{ \bd{e} }^{(2)}_i (t)   +
 \tilde{S}^{\parallel}_i \hat{\bd{m}}_i  (t).
 \label{eq:tildeSexpand}
 \end{equation}
Evaluating the time derivative in Eq.~(\ref{eq:HBdef}) with the help
of the formula (\ref{eq:Fey}) and inserting the expansion (\ref{eq:tildeSexpand}) 
for the rotated spin operators we can rewrite the Berry-phase contribution to the
effective Hamiltonian as
 \begin{equation}
  \tilde{{\cal{H}}}_B ( t )  =  - \sum_i  \Bigl[ \omega_i^{(1)} ( t ) \tilde{S}^{(1)}_i 
 + \omega_i^{(2)} ( t ) \tilde{S}^{(2)}_i + \omega_i^{\parallel} ( t ) \tilde{S}^{\parallel}_i
 \Bigr],
 \label{eq:HBEuler}
 \end{equation}
where the three time-dependent energies
 $\omega_i^{(1)} ( t )$, $\omega_i^{(2)} ( t )$, and $\omega_i^{\parallel} ( t )$
can be identified with the well known 
Euler angle parametrization of 
the components of the angular velocity
vector in the rotating reference frame:\cite{McCauley97} 
 \begin{subequations} 
\begin{eqnarray}
 \omega_i^{(1)} ( t ) & = &  
  \dot{\varphi}_i \sin \theta_i \sin \psi_i + \dot{\theta}_i \cos \psi_i,
 \\
 \omega_i^{(2)} ( t ) & = &
  \dot{\varphi}_i \sin \theta_i \cos \psi_i - \dot{\theta}_i \sin \psi_i ,
 \\
\omega_i^{\parallel} ( t ) & = &
 \dot{\varphi}_i \cos \theta_i +\dot{\psi_i}.
 \end{eqnarray}
\end{subequations}
In the models discussed in this work  the proper rotation of the comoving basis 
is irrelevant, so that we may focus on the special case  $\bd{\psi}_i ( t ) =0$.
The Berry-phase Hamiltonian (\ref{eq:HBEuler}) then reduces to
\begin{align}
 \tilde{{\cal{H}}}_B ( t )  =  - \sum_i  \Bigl[ &
   \dot{\theta}_i  \tilde{S}^{(1)}_i +  \dot{\varphi}_i \sin \theta_i \tilde{S}^{(2)}_i
  + \dot{\varphi}_i \cos \theta_i \tilde{S}^{\parallel}_i
 \Bigr].
 \label{eq:dotHfinal}
 \end{align}

\begin{figure}[tb]    
   \centering
  \includegraphics[width=\linewidth]{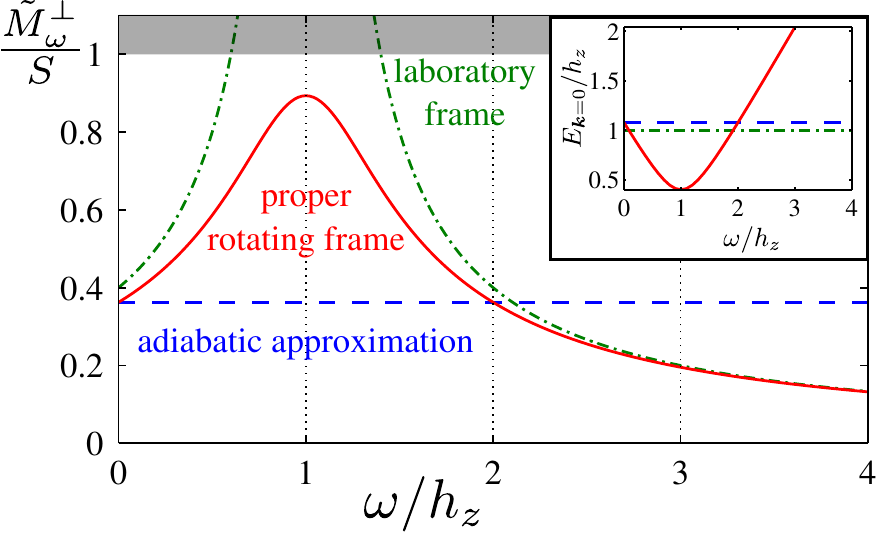}\nspace\nspace
  \caption{%
(Color online)
Comparison of the results for the perpendicular magnetization $\tilde{M}_\omega^\perp$ as a function
of the rotation frequency $\omega$ in the three different approaches for our
model system given in Eq.~(\ref{eq:Hdefh}) with the parameters
\mbox{$J=h_z$}, \mbox{$h_\perp=0.4h_z$}, $S=1/2$ at the temperature \mbox{$T=0.5h_z$}:
blue (dashed) line presents the results of the adiabatic approximation, Eq.~(\ref{eq:Mad}),
green (dash-dotted) line is perturbation theory in the laboratory frame, Eq.~(\ref{eq:Mlab})
and red (solid) is the proper rotating frame.
The shaded area indicates the unphysical region. The inset shows a sketch of the
spin-wave gap $E_{\bd k=0}/h_z$ as a function of rotation frequency for the same parameters.
}
    \label{fig:comparison}
  \end{figure}

For the  rotating ferromagnet shown in Fig.~\ref{fig:heisrot},
symmetry suggests that the proper rotating coordinate system 
is characterized by a time-dependent  
precession angle $\varphi_i ( t ) = - \omega t$ 
and a constant nutation angle $\theta$. 
The Berry-phase contribution (\ref{eq:dotHfinal}) to the Hamiltonian
in the rotating basis is then
\begin{equation}
 \tilde{{\cal{H}}}_B = \omega \sum_i [ \sin \theta \tilde{S}^{(2)}_i + \cos \theta 
 \tilde{S}^{\parallel}_i
 ]\;\label{eq:Hberr},
\end{equation}
which is independent of time.
Next we express the spin components
in the rotating reference frame in terms of a third type of 
Holstein-Primakoff boson $c_i$, which should not be confused
with the Holstein-Primakoff boson $a_i$  introduced 
in the adiabatic basis, and also not 
with the laboratory basis Holstein-Primakoff boson $b_i$.
The  true tilt angle $\theta$ is determined from the requirement that
the effective Hamiltonian contains no terms linear in the bosons,
which yields the frequency-dependent result
\begin{equation}
\cos\theta  =  \frac{h_z - \omega} { \tilde{h}_{\omega}}
\end{equation}
where 
 \begin{equation}
\tilde{h}_{\omega} =[ h_{\bot}^2 + (h_z - \omega )^2 ]^{1/2}.
 \end{equation}
Note that for finite $\omega$ the true tilt angle $\theta$ is larger
than the angle $\theta_0$ between rotation axis and magnetic field.
In fact,
our result for $\theta$ agrees with the result for
a single isolated spin in a rotating magnetic 
field given in the book by Bohm {\it{et~al.}}~\cite{Bohm03}
For the specific geometry shown in Fig.~\ref{fig:heisrot} the
proper rotating reference frame has also been discussed
previously in Ref.~[\onlinecite{Lin05}], 
but our Eq.~(\ref{eq:dotHfinal}) is more general.
In fact, our many-body approach 
allows us to set up a systematic $1/S$ expansion and calculate the
thermodynamics and the
correlation functions of any time-dependent spin model with
finite local moments.
Following the steps of the spin-wave expansion we obtain the
Hamiltonian
\begin{equation}
 {{\cal H}} \approx \sum_{\bd k} E_{\bd k} c_{\bd k}^\dagger c_{\bd k}
\end{equation}
to quadratic order in the bosonic operators $c_{\bd k}^\dagger$ and $c_{\bd k}$
describing bosons in the proper rotating reference frame.
The dispersion $E_{\bd k}=\epsilon_{\bd k}+\tilde{h}_\omega$
is modified by the finite oscillation frequency, see inset in Fig.~\ref{fig:comparison}. Imposing suitable
initial conditions for our model, we obtain for the
time-dependent magnetization in linear spin-wave theory,
 \begin{align}
 \bd{M} (t )&
= \tilde{M}_{\omega}^\parallel \hat{\bd{z}}+\tilde{M}_{\omega}^\perp [   \cos ( \omega t )  \hat{\bd{x}} -
\sin ( \omega t )  \hat{\bd{y}}  ]\notag\\
&=  \hat{\bd{m}}_{\omega} ( t ) \tilde{M}_{\omega} \;,
 \label{eq:Mres}
\end{align}
where $\tilde{M}_{\omega}^\parallel=\cos\theta  \tilde{M}_{\omega}$ and $\tilde{M}_{\omega}^\perp=\sin\theta  \tilde{M}_{\omega}$ with
\begin{equation}
 \tilde{M}_{\omega}  = S - \frac{1}{N} \sum_{\bd{k}} \frac{1}{e^{ \beta E_{\bd k}  } -1 }\;,
 \end{equation}
and 
$\hat{\bd{m}}_{\omega} ( t ) =    \sin \theta [  \cos ( \omega t )  \hat{\bd{x}} -
\sin ( \omega t )  \hat{\bd{y}}  ] + \cos \theta  \hat{\bd{z}}$.
In the limit $\omega \rightarrow 0$, Eq.~(\ref{eq:Mres}) reduces to the
result (\ref{eq:Mad}) of the adiabatic approximation, which is only accurate as long as
$ | \omega | \ll h$. In fact, for the two special cases  $\omega=0$ and $\omega=2h_z$ where the
effective field is equal to the external field $h=\tilde h_\omega$,
the adiabatic approximation Eq.~(\ref{eq:Mres}) matches the correct result of Eq.~(\ref{eq:Mad}).
While the  result (\ref{eq:Mlab}) for the magnetization
obtained from perturbation theory in the laboratory basis
approaches the more accurate rotating reference frame result
(\ref{eq:Mres})
 for $\omega \rightarrow 0$ and for large frequencies $\omega \gtrsim 2 h_z$,
perturbation theory in the laboratory basis
gives unphysical results in the vicinity of the resonance $ | h_z - \omega | \lesssim h_{\bot}$
and is thus meaningless, whereas Eq.~(\ref{eq:Mres}) predicts that
the magnetization simply rotates in the $xy$-plane ($\theta \approx \pi /2$).
Note that the magnetization  shown in Fig.~\ref{fig:comparison} does not 
approach $\tilde{M}_\omega^\perp=S$
because thermal fluctuations suppress the total magnetic moment.
 
\sect{Parametric resonance and BEC of magnons in 
YIG}
\label{sec:BEC_YIG}
Next, let us study another time-dependent spin model which
gives us some insight into the 
relation between parametric resonance, BEC of magnons, and the reorganization
of the magnetic state.
Previously, this problem has been 
addressed in Refs.~[\onlinecite{Zvyagin85,Zvyagin07}]
using a Heisenberg ferromagnet with static single-ion anisotropy
in a time-dependent magnetic field.
For our purpose it is more convenient to 
consider a modified version of this model, involving
a static magnetic field in $z$ direction 
and a rotating single-ion anisotropy of magnitude $A$,
\begin{align}
 {\cal{H}} ( t ) &=   -\frac 12 \sum_{ij} J_{ij} \bd S_i \cdot \bd S_j 
 - h   \sum_i S_i^z   
 \nonumber
 \\
 &
 - \frac{A}{2}\sum_i \left\{  [ \bd{S}_i \cdot \bd{n} ( t ) ]^2
 - [ \bd{S}_i \cdot ( \hat{\bd{z}} \times \bd{n} ( t )) ]^2 \right\},
 \label{eq:Hion}
 \end{align}
\begin{figure}[tb]    
   \centering
  \includegraphics[width=0.384\textwidth]{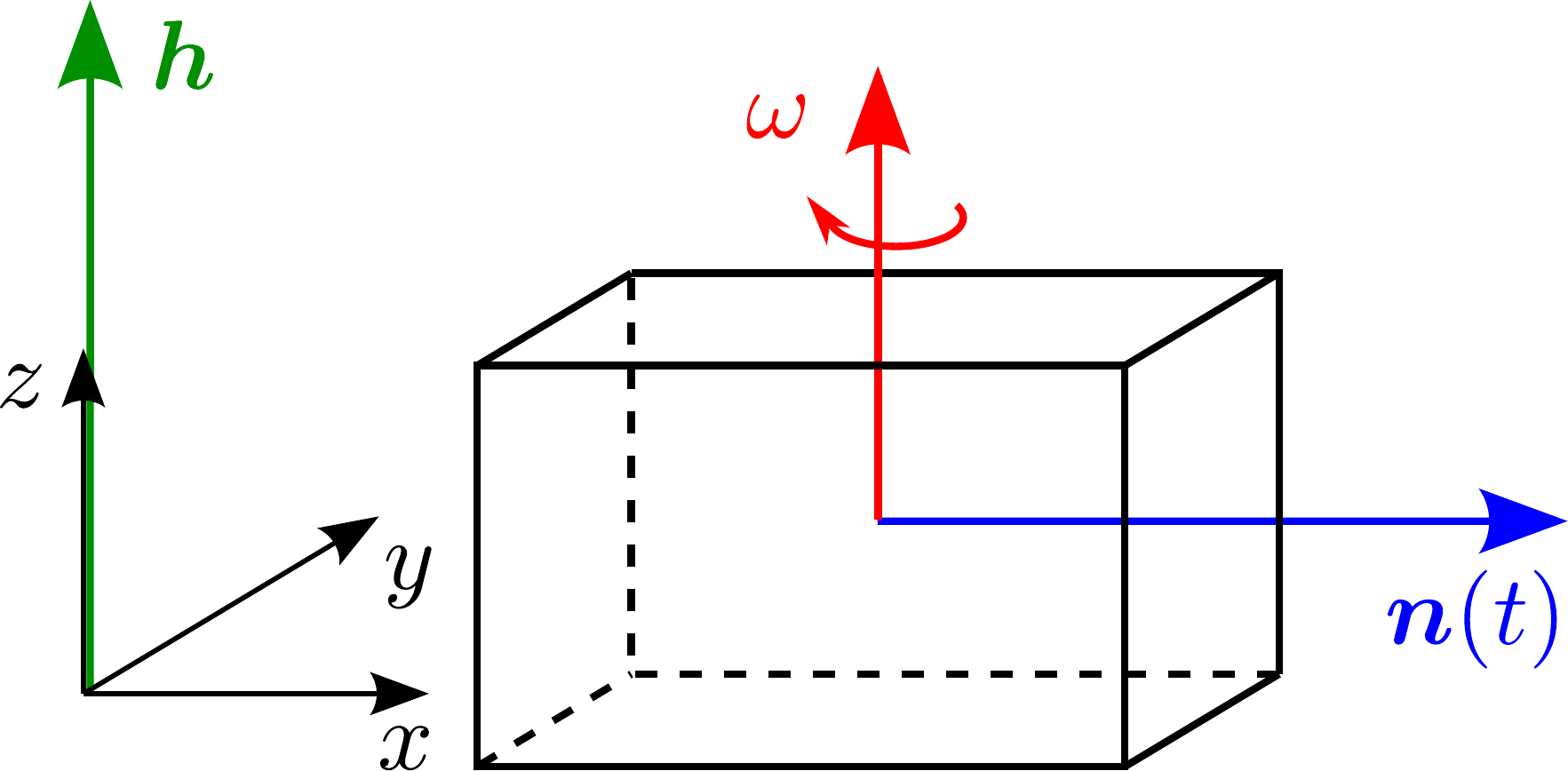}\nspace
  \caption{%
(Color online)
Graph of the time dependent spin model defined in Eq.~(\ref{eq:Hion}).
The model can be interpreted as 
a  Heisenberg magnet with a single-ion  anisotropy axis
$\hat{\bd{n}}$ that is fixed in the laboratory frame.
The magnet is exposed  to a static external field  
perpendicular to the anisotropy axis and rotates counter-clockwise 
around an axis parallel to the field.
}
    \label{fig:singleion}
  \end{figure}where the anisotropy axis
$\bd{n} ( t ) = \cos ( \omega t ) \hat{\bd{x}} - \sin ( \omega t ) \hat{\bd{y}}$ 
rotates clockwise in the $xy$ plane.
An illustration of the model (\ref{eq:Hion}) is shown in Fig.~\ref{fig:singleion}.
After bosonization of the spins using
the Holstein-Primakoff transformation in the laboratory basis,
we obtain in linear spin-wave theory, 
 \begin{equation}
 {\cal{H}} ( t )  \approx  
  \sum_{\bd{k}} 
 \Bigl[ ( \epsilon_{\bd{k}} + h )  b^{\dagger}_{\bd{k}} b_{\bd{k}}
 + \frac{h_c}{2}  (   e^{ 2 i \omega t }  b_{ - \bd{k}} b_{\bd{k}}
 + \mbox{H.c.} ) 
 \Bigr],
 \label{eq:H2iondyn}
 \end{equation}
with $h_c = A S$.  Time-dependent boson models of this form have been
studied as model systems for parametric resonance in
magnon gases. \cite{Akhiezer68,Schloemann63,Zakharov70} 
In fact, with appropriate replacements \cite{footnoteYIG} the
magnon Hamiltonian for YIG in an external microwave field
parallel to the external field
has the same form as Eq.~(\ref{eq:H2iondyn}).
It is well known~\cite{Zakharov70} that 
the Hamiltonian
(\ref{eq:H2iondyn})  predicts a parametric instability
of the magnons with 
 wave-vectors in the regime 
 $
 h_c > | \epsilon_{\bd{k}} + h - \omega |$. 
If this condition is satisfied, then the magnon occupation
grows exponentially during some intermediate time interval, until it
saturates and the
system approaches a new equilibrium state, which in principle can be
calculated by taking the interactions between the magnons into account.
Here we show that the 
dynamics of the local magnetization
$\langle {\bd{S}}_i ( t ) \rangle$ as well as the  magnon spectrum
can be obtained using our  time-dependent spin-wave formalism
\textit{without considering interactions between magnons}.
Because the Hamiltonian  (\ref{eq:H2iondyn})
has the same spin symmetries as the
rotating ferromagnet discussed above,
the proper rotating reference frame 
is again given by a time-dependent precession angle $\varphi_i ( t ) = - \omega t$ and
a constant nutation angle $\theta$.
The  Berry-phase Hamiltonian $ \tilde{{\cal{H}}}_B = \omega \sum_i [ \sin \theta \tilde{S}^{(2)}_i + \cos \theta 
 \tilde{S}^{\parallel}_i
 ]$ is therefore identical with 
the rotating ferromagnet discussed above, see Eq.~(\ref{eq:Hberr}).
It is then easy to show that for 
$|h - \omega | > h_c$ all spins point in the direction of the field so that
the tilt angle $\theta$ vanishes and
the magnon spectrum is
  \begin{figure}[tb]    
   \centering
  \includegraphics[width=.8\linewidth]{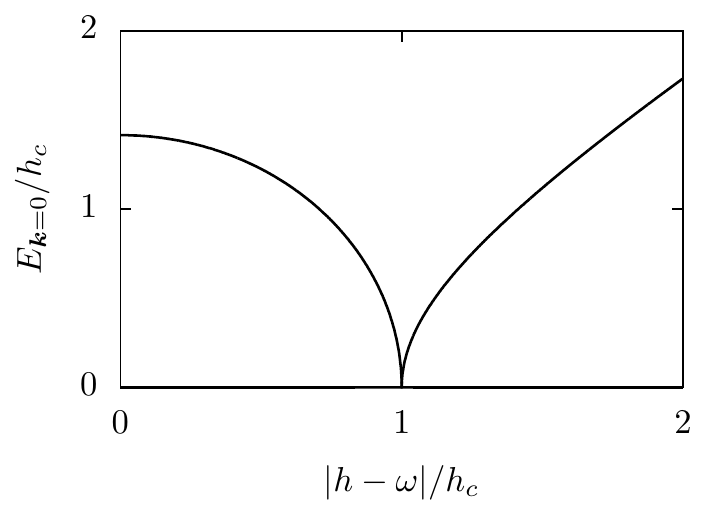}\nspace\nspace
  \caption{%
Plot of the spin-wave gap $E_{\bd{k}=0}$ 
of the time-dependent spin model defined in Eq. (\ref{eq:Hion}) as a function of $|h - \omega |/h_c$. 
}
    \label{fig:energygap}
  \end{figure}
\begin{equation}
E_{\bd{k}} = \sqrt{ ( \epsilon_{\bd{k}} + h - \omega)^2 - h_c^2 }\;,
\end{equation}
where again $\epsilon_{\bd{k}} = S ( J_0 - J_{\bd{k}} )$.
On the other hand,  
for $ | h - \omega | < h_c$ the angle between magnetic field and
magnetization does not vanish,
\begin{equation}
\cos \theta = \frac{ h - \omega }{h_c}\;,
\end{equation}
 and
the magnon spectrum is
\begin{equation}
 E_{\bd{k}}^2 = 
 \sqrt{\Bigl[ \epsilon_{\bd{k}} +     \frac{3h_c}{2}   - \frac{( h - \omega)^2}{ 2 h_c} 
 \Bigr]^2 -     \Bigl[  \frac{h_c}{2} + \frac{(h - \omega )^2}{2h_c}
 \Bigr]^2}  .
 \label{eq:EkAomega}
 \end{equation}
A graph of the spin-wave gap $E_{\bd k=0}$ is presented in Fig.~\ref{fig:energygap}.
In the tilted phase, the time-dependent magnetization is in linear spin-wave theory
$\bd{M} ( t ) = \tilde{M}_{\omega} \hat{\bd{m}}_{\omega} ( t )$,
where $\hat{\bd{m}}_{\omega} ( t ) =    \sin \theta [  \cos ( \omega t )  \hat{\bd{x}} -
\sin ( \omega t )  \hat{\bd{y}}  ] + \cos \theta  \hat{\bd{z}}$ and
 \begin{align}
 \tilde{M}_{\omega} & =  S + \frac{1}{2}  
  - \frac{1}{N} \sum_{\bd{k}} \frac{1}{  E_{\bd{k}} } \Bigl[  \epsilon_{\bd{k}} + \frac{3h_c}{2} - \frac{( h - \omega)^2}{ 2 h_c} \Bigr] \nonumber\\
  &  \hspace{2.8cm} \times 
\Bigl[ \frac{1}{e^{ \beta E_{\bd{k}}  } -1 } + \frac{1}{2} \Bigr].
 \label{eq:tildeMdef3}
 \end{align}
Note that the gap $E_{\bd{k}=0}$ of the magnon energy vanishes at the
critical fields $h_c^{\pm} = \pm h_c + \omega$, signaling
a quantum phase transition. Because the magnetic state in the tilted phase
spontaneously breaks the $Z_2$-symmetry $\bd{S}_i\cdot \bd{n} \rightarrow - \bd{S}_i\cdot \bd{n}$
of the spin Hamiltonian (\ref{eq:Hion}), 
this phase transition belongs to the Ising universality class.
If we bosonize the spin operators
in the laboratory frame, then at the critical point the corresponding 
bosons acquire a macroscopic expectation value, 
which corresponds to BEC of magnons.~\cite{Matsubara56,Batyev84}
However, 
as pointed out by Kohn and Sherrington,~\cite{Kohn70} such a transition is neither  
accompanied by magnon superfluidity nor by
off-diagonal long-range order, which distinguishes
the magnon condensate from the BEC of trapped  atoms or molecules.
In fact, the macroscopic occupation of magnon modes is an
artifact of working in the laboratory frame;
the magnons defined in the proper rotating reference frame
never condense.

\begin{figure}[tb]    
   \centering
  \includegraphics[width=0.87\linewidth]{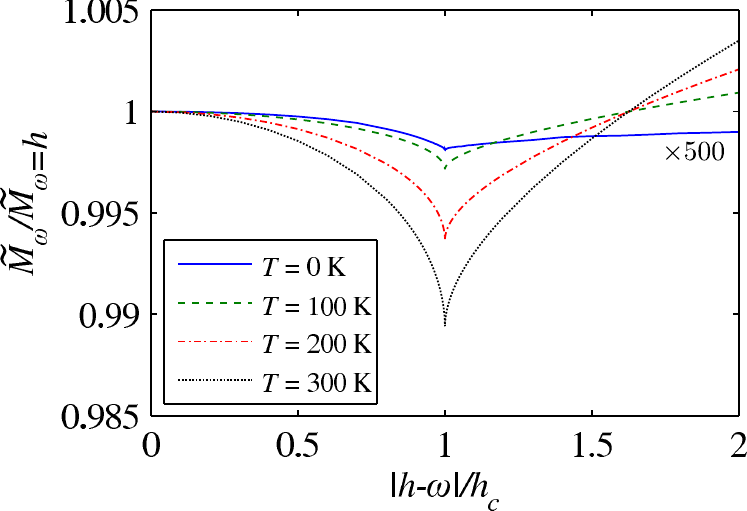}\nspace
  \caption{(Color online)
Length $\tilde{M}_{\omega}$ of the magnetization vector 
defined in Eq.~(\ref{eq:EkAomega})
at different temperatures $T$
for typical parameters  describing the pumped magnon gas in bulk
YIG \cite{footnoteYIG} ($J=1.29~\mathrm{ K}$, $h_c =0.55~\mathrm{ K}$, $S=14.2$). 
The Curie temperature of YIG is $T_c = 560~\mathrm{ K}$. \cite{Cherepanov93,Melkov96} 
Note that for $T < T_c$  the magnetization in three dimensions is finite
for all values of the reduced magnetic field $h-\omega$.
The deviation of the $T=0~\mathrm{ K}$ result from unity is 
enhanced by a factor of 500.
}
    \label{fig:Momega}
  \end{figure}

Given the fact that our model Hamiltonian (\ref{eq:Hion}) has the same symmetries
as the effective spin Hamiltonian for YIG,~\cite{Cherepanov93}
with appropriate substitutions \cite{footnoteYIG} 
our model can be used to understand
the nonequilibrium dynamics of the magnetization in YIG 
in the vicinity of the condensation transition.
In Fig.~\ref{fig:Momega}, we show a numerical evaluation of
the frequency-dependent
magnetization  $\tilde{M}_{\omega}$ given in Eq.~(\ref{eq:tildeMdef3})
using effective parameters for YIG.~\cite{footnoteYIG}
We predict that close to the threshold of BEC
the magnetization  shows a characteristic
dip  of the order of $1 \%$ at relevant temperatures.

\sect{Conclusions}
\label{sec:conclusions}

In summary, we have developed a general method to
set up the spin-wave expansion for  time-dependent spin models.
Our method is very general and should also be useful to study 
nonequilibrium phenomena in all kinds of ordered magnets, including
quantum antiferromagnets and frustrated magnets with finite local moments.
We have used our method to study a simplified spin model
for the magnon gas in YIG, and have
shown that magnon BEC in this system can be interpreted as a 
magnetic quantum phase transition belonging to the Ising universality class.
Our prediction of a dip in the magnetization
close to the threshold for BEC can be tested experimentally.

We thank M. Taillefumier, V. Vasyuchka, and A. Serga for discussions.
This work was financially supported by the DFG via SFB/TRR 49.
\nspace


\begin{thebibliography}{99}
\bibitem{footnoterot}
We use the so-called Rodrigues formula 
  $ e^{ \bd{\varphi} \times } \bd{h}   
= \hat{\bd{\varphi}} ( \hat{\bd{\varphi}} \cdot \bd{h} )
 +   \sin \varphi   \,   \hat{\bd{\varphi}} \times \bd{h}  
 -   \cos \varphi   \,  \hat{\bd{\varphi}} \times  (  \hat{\bd{\varphi}} \times \bd{h} ) 
 $
to represent the rotation of a vector $\bd{h}$ 
around an axis
 $\hat{\bd{\varphi} } = \bd{\varphi} / | \bd{\varphi } |$
with angle $ \varphi = | \bd{\varphi }|$
in terms of an exponentiated cross product.
\end{thebibliography}
\end{document}